\documentstyle[psfig]{l-school} 
\begin{document}
\markboth{H.C. Pauli}{DERIVING THE EFFECTIVE INTERACTION}
\setcounter{part}{12}                  
\title{On Deriving the Effective Interaction\\
from the QCD-Lagrangian}
\author{H. C. Pauli}
\institute{Max-Planck-Institut f\"ur Kernphysik \\
                   D-69029 Heidelberg}
%
\maketitle
%
The canonical front form Hamiltonian for non-Abelian  SU(N)
gauge theory in 3+1 dimensions is mapped
non-perturbatively 
on an effective  Hamiltonian which acts 
only in the Fock space of a quark and an antiquark.
The approach is based on the novel method of iterated
resolvents and on discretized light-cone quantization, 
driven to the continuum limit. It is free of the usual
Tamm-Dancoff truncations of the Fock space, rather
the perturbative series are consistently resumed to all orders
in the coupling constant. 
Emphasis is put on  dealing with the many-body
aspects of gauge field theory. 
Important is that the higher Fock-space amplitudes 
can be retrieved self-consistently  from these solutions.
 
\section{THE STRUCTURE OF THE HAMILTONIAN}
\label{sec:2} 

In canonical field theory the  four  components of
the  energy-momentum vector $ P ^\nu$ commute and 
are constants of the motion. 
In the front form of Hamiltonian dynamics \cite{dir49} 
they are denoted \cite{leb80} by 
$ P ^\nu = (P  ^+, \vec P  _{\!\bot}, P  ^-)$.
Its spatial components $\vec P  _{\!\bot}$  and $P  ^+$ 
are independent of the interaction and diagonal in
momentum representation.
Their eigenvalues are the sums of the single particle
momenta, 
$     P ^+ = \sum p^+ $ and
$     \vec P  _{\!\bot} = \sum\vec p _{\!\bot}$. 
Each single particle has four-momentum 
$ p^\mu = (p^+, \vec p _{\!\bot}, p^- )$ and 
sits on its mass-shell $ p^\mu p_\mu = m^2$. 
Each particle state ``$q$''  is then characterized by six 
quantum numbers
$     q = (p^+, \vec p_{\!\bot}, \lambda;c, f ) $: 
the three spatial momenta, helicity, color and flavor.
The temporal component $P ^-= 2 P_+ $ depends on the
interaction  and is a complicated non-diagonal
operator \cite{brp91,bpp97}. It propagates the system in the 
light-cone time $ x^+ = x^0 + x^3$.
The contraction of  $P ^\mu$  is the operator of
invariant mass-squared,
\begin {equation} 
      P ^\mu P _\mu  = P ^+ P ^- - \vec P  _{\!\bot} ^{\,2} 
      \qquad  \equiv H _{\rm LC} \equiv H
\ .\end {equation} 
It is Lorentz invariant and  refered to 
somewhat improperly but conveniently 
as the `light-cone Hamiltonian' 
$ H _{\rm LC} $, or shortly $ H $. 
One seeks a representation in which $H$ is diagonal
$ H \vert \Psi\rangle = E\vert \Psi\rangle $.
Introductory texts \cite{brp91,gla95,bpp97} are available. 

The (light-cone) Hamiltonian is split for convenience into 
four terms:
$      H = T + V + F + S$.
The kinetic energy  $T$ is diagonal in Fock-space 
representation,  its eigenvalue is the {free invariant mass squared}
of the particular Fock state. 
The vertex interaction $ V $ is the relativistic interaction
{\it per se}. It has  no diagonal matrix elements, 
is linear in $g$ and changes the particle number 
by 1. The instantaneous interactions $ F $ and $ S $ are 
consequences of working in the light-cone gauge  $A^+=0$.
They are proportional to $g^2$.  
By definition, the seagull interaction $S$  conserves
the particle number and the fork interaction $F$ changes
it by 2. 
As a consequence of working in light-cone representation 
and in the light cone gauge, the vacuum state has no matrix
elemens with any of the Fock states:
The vacuum {\em does not fluctuate}.

\begin{figure} 
\psfig{figure=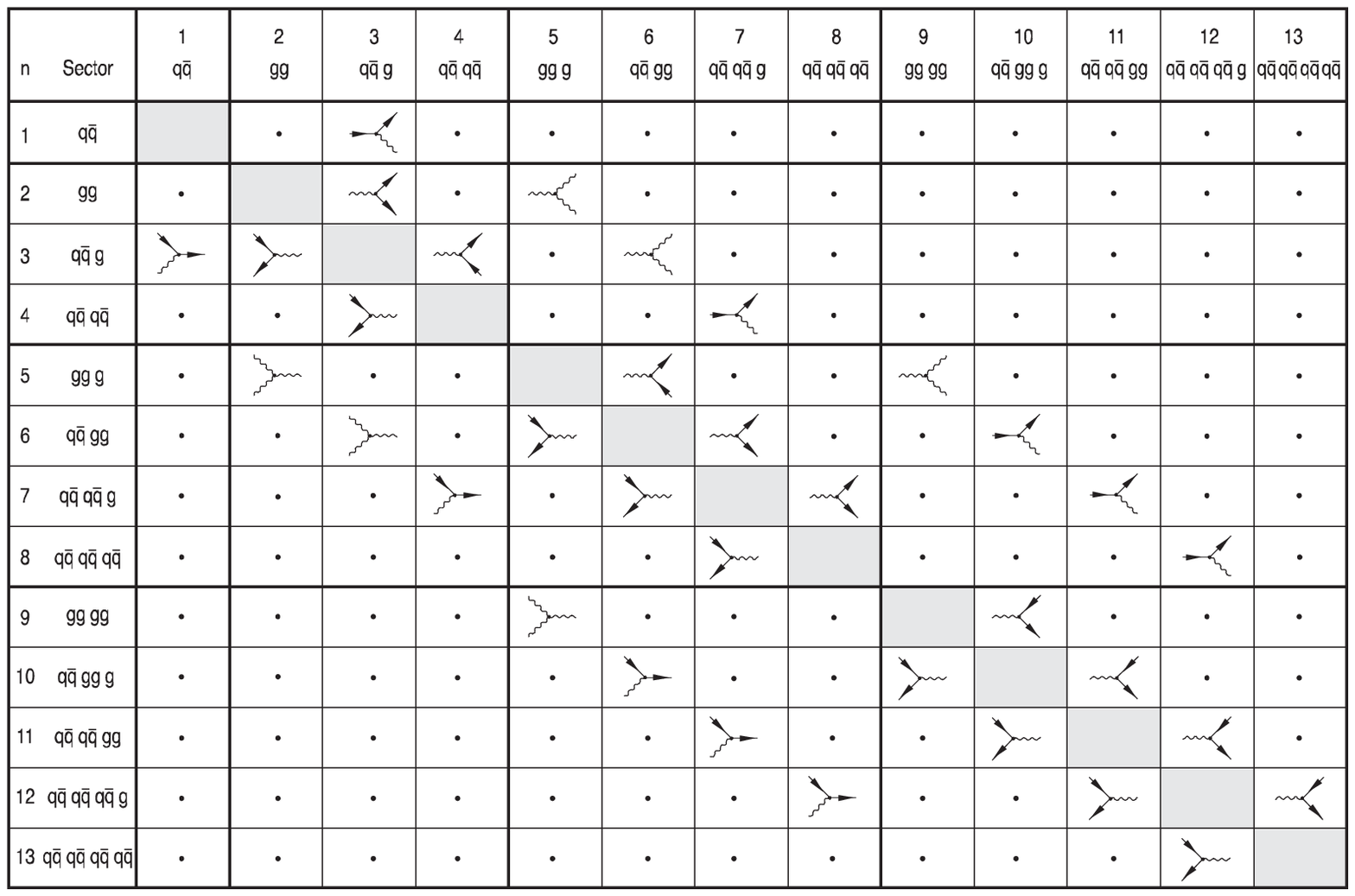,width=120mm} 
\caption{\label{fig:holy-2} 
    The Hamiltonian matrix for a SU(N)-meson and 
     a harmonic resolution $K=4$.
     Only vertex diagrams are included.
     Zero matrices are marked by a dot ($\cdot$). 
} \end{figure} 

The eigenvalue equation $H\vert\Psi\rangle=E\vert\Psi\rangle$ 
stands for an infinite set of coupled integral equations \cite{bpp97}.
The eigenfunctions $\vert\Psi\rangle$ are superpositions of all
Fock-space projections like $\langle q\bar q\vert\Psi\rangle$ or 
$\langle q\bar q\,g\vert\Psi\rangle$. 
However, if one works with discretized light-cone quantization 
(DLCQ \cite{pab85a}), one deals with a {\em finite set} of 
coupled matrix equations.
As consequence of periodic  boundary conditions the single 
particle momenta are discrete and the Fock states are 
denumerable and orthonormal. The Hamiltonian matrix is
illustrated in Fig.~\ref{fig:holy-2} for the 
{\em harmonic resolution} $K\equiv P^+L/(2\pi)=4$.
Its rows and columns are the Fock states which are grouped
into the Fock-space sectors $q\bar q$, $q\bar q\,g, \dots\ $, 
denumerated for simplicity by $n=1,2,\dots,13$. 
Within each sector one has many Fock states, which differ
by the helicities, colors, and momenta of the single particles,
subject to fixed total momenta $P ^+$ and $\vec P_{\!\bot}$. 
Since $P  ^+$ has only positive eigenvalues and since each 
particle has a lowest possible value of $p^+$, the number of 
particles in a Fock state is limited for any fixed value of $K$:
The number of Fock space sectors is {\em finite}. 
The transversal momenta $\vec p_{\!\bot}$ can take either sign. 
Their number must be regulated by some convenient cutoff,
(Fock-space regularization, {\it f.e.} see \cite{leb80}).
The most remarkable property of the Hamiltonian matrix
is its block structure and its sparseness. Most of the blocks
are zero matrices, marked by a dot ($\cdot$) in the figure.
Only those blocks in Fig.~\ref{fig:holy-2} denoted by the 
graphical symbol of a vertex are potentially non-zero, with the 
actual matrix elements tabulated elsewhere \cite{brp91,bpp97}.

Here then is the problem, the bottle neck of any Hamiltonian
approach in  field theory: The dimension of the Hamiltonian
matrix  increases exponentially fast.
Suppose, the regularization procedure allows for 10 discrete
momentum states in each direction. A single particle has then 
about $10^3$ degrees of freedom.
A Fock-space sector with $n$ particles
has then roughly  $10^{n-1}$ different Fock states.
Sector 13 with its  8 particles has thus about  $10^{21}$
and sector 1 ($q\bar q$) about $10^3$ Fock states. 
What one wishes instead of is to derive an effective interaction
which acts only in the comparatively small $q\bar q$-space,
like a Coulomb interaction acts only between an electron and a 
positron. Loosely speaking, the aim of
deriving an effective interaction can be understood as 
reducing the dimension in a matrix diagonalization problem 
from $10^{21}$ to $10^{3}$!

The first attempt to formulate an effective interaction in field theory
is the approach of Tamm and Dancoff (TDA \cite{tam45,dan50}).
The present `method of iterated resolvents' \cite{pau93,pau96}
is closely related to TDA.
Alternatively, one can apply a sequence of analytical
and approximate unitary transformations in order to  render the 
Hamiltonian matrix more and more `band diagonal'. Two
methods have been proposed recently,
the `similarity transform' of  Glazek and Wilson, see \cite{wwh94}, 
and the `Hamiltonian flow equations' of Wegner \cite{weg94}.
All of these methods are under active research. 
Applications particularly by 
Trittmann {\it et al} \cite{trp97,trp97a,tri97}, 
Brisudova {\it et al} \cite{brp95,bpw97},
Jones {\it et al} \cite{jop96a,jop96b},
and Gubankova {\it et al} \cite{guw97}
are presented elsewhere in these proceedings.
Common to most of them is that one truncates the Fock space 
before approximation methods are applied. 
No truncation of the Fock space is needed with the method of 
iterated resolvents, as to be shown below. 

\section {THE METHOD OF ITERATED RESOLVENTS} 
\label {sec:3}

The method of effective interactions is a well known tool in  
many-body physics which in field theory is known as the 
Tamm-Dancoff-approach \cite{tam45,dan50}. The method is 
ideally suited for a field theory, because Fock-space sectors 
$\vert i \rangle$ appear in the most natural way.
The Hamiltonian matrix can be understood as a matrix of block 
matrices, whose rows and columns are enumerated by
$i=1,2,\dots N$ like in Fig.~\ref{fig:holy-2}.
The eigenvalue equation can be written as a coupled set of 
block matrix equations:
\begin {equation} 
      \sum _{j=1} ^{N} 
      \ \langle i \vert H \vert j \rangle 
      \ \langle i \vert \Psi\rangle 
      = E\ \langle n \vert \Psi\rangle 
\, \qquad {\rm for\ all\ } i = 1,2,\dots,N 
\ .\label {eq:319}\end {equation} 
In TDA, the rows and columns are split into a $P$-space,
$P=\vert1\rangle\langle1\vert$, and the rest, the $Q$-space. 
Explicitly written out, the eigenvalue
equation (\ref{eq:319}) becomes a set of two coupled block matrix 
equations: 
\begin {eqnarray} 
   \langle P \vert H \vert P \rangle\ \langle P \vert\Psi\rangle 
 + \langle P \vert H \vert Q \rangle\ \langle Q \vert\Psi\rangle 
 &=& E \:\langle P \vert \Psi \rangle 
 \ ,  \label{eq:321} \\   {\rm and} \qquad
   \langle Q \vert H \vert P \rangle\ \langle P \vert\Psi\rangle 
 + \langle Q \vert H \vert Q \rangle\ \langle Q \vert\Psi\rangle 
 &=& E \:\langle Q \vert \Psi \rangle 
\ . \label{eq:322}\end {eqnarray}
Rewrite the second equation as
$      \langle Q \vert E  -  H \vert Q \rangle 
    \ \langle Q \vert \Psi \rangle 
  =   \langle Q \vert H \vert P \rangle 
    \ \langle P \vert\Psi\rangle $,
and observe that the quadratic matrix 
$ \langle Q\vert E -  H \vert Q \rangle $ could be inverted to 
express the Q-space wave-function $\langle Q\vert\Psi\rangle$
in terms of $\langle P \vert\Psi\rangle$. But the eigenvalue $E$ 
is unknown at this point.  One introduces therefore a redundant 
parameter $\omega$, and defines $G _ Q (\omega)   =  \left[
    \langle Q \vert\omega- H \vert Q \rangle\right]^{-1} $.
The so obtained effective interaction
\begin{equation} 
      \langle P \vert H _{\rm eff} (\omega) \vert P \rangle =
      \langle P \vert H \vert P \rangle +
      \langle P \vert H \vert Q \rangle 
      \ G _ Q (\omega) \ %
      \langle Q \vert H \vert P \rangle 
\ .  \label{eq:340}\end{equation} 
acts only in the much smaller $P$-space:
$    H _{\rm eff} (\omega ) \vert\Phi_k(\omega)\rangle =
      E _k (\omega )\,\vert\Phi _k (\omega ) \rangle $.
Varying $\omega$ one generates a set of {energy functions} 
$E _k(\omega) $. All solutions of the fix-point equation 
$E _k (\omega ) = \omega $ are eigenvalues of the full
Hamiltonian $H$. In fact one can find all eigenvalues of $H$
despite the fact that the dimension of $H _{\rm eff}$ is usually
much smaller than that of the full $H$ \cite{pau93,pau96}.
The procedure is formal, since the inversion of the
Q-space matrix is as complicated as its diagonalization.
The adventage is that resolvents can be approximated 
systematically:
The two resolvents
\begin {equation}
     G _Q (\omega) =  {1\over \langle Q \vert 
           \omega - T - U  \vert Q \rangle} 
     \quad{\rm and}\quad 
     G_Q ^{(0)}  (\omega) =  {1\over \langle Q \vert 
           \omega - T \vert Q \rangle} 
\ , \label {eq:352} \end {equation} 
defined once with and once without the non-diagonal 
interaction $U$, 
are identically related by 
$G_Q (\omega) = G_Q^{(0)}(\omega)
+G_Q^{(0)}(\omega)\,U\,G_Q(\omega)$, 
or by the infinite series of perturbation theory.
Albeit exact in principle, the Tamm-Dancoff-approach (TDA) 
suffers from a practical aspect:  The approach
is useful only if one truncates the perturbative 
series to its very first term. This destroys
Lorentz and gauge invariance, and requires a sufficiently
small coupling constant. 

Truncation can be avoided by `going backwards'.
Reinterpreting the $Q$-space as the last sector $N$
(sector 13 in Fig.~\ref{fig:holy-2}) 
and the $P$-space with the rest, 
the above equations can be interpreted as to reduce the block 
matrix dimension from $N$ to $N-1$, with an effective interaction
acting now in the smaller space of $N-1$ blocks. 
This procedure can be iterated, from $N-1$ to $N-2$, and so
on, until one arrives at block matrix dimension 1 
where the procedure stops: 
The effective interaction in the Fock-space
sector with only one quark and one antiquark is defined
unambiguously.
One has to deal then with `resolvents of resolvents', 
or with iterated resolvents \cite{pau96}. 

To be more explicit suppose one has arrived in the course of 
this reduction at block matrix dimension $n$, with 
$1\leq n\leq N$.  Denote the corresponding effective 
interaction  $H_n (\omega)$. The eigenvalue problem reads then
\begin {equation} 
   \sum _{j=1} ^{n} \langle i \vert H _n (\omega)\vert j \rangle 
                      \langle j \vert\Psi  (\omega)\rangle 
   =  E (\omega)\ \langle i \vert\Psi (\omega)\rangle 
\ , \ \quad{\rm for}\ i=1,2,\dots,n
\,. \label {eq:406} \end {equation}
Observe that $i$ and $j$ refer here to sector numbers, 
and that n refers to both, the last sector number, 
and the  number of sectors.
Now,  like in the above, define the resolvent as the inverse 
of the Hamiltonian in the last sector
\begin {eqnarray} 
            G _ n (\omega)   
&=&   {1\over \langle n \vert\omega- H_n (\omega)\vert n \rangle} 
\, \label {eq:410} \\  {\rm thus}\qquad  
            \langle n \vert \Psi (\omega)\rangle   
&=&   G _ n (\omega) 
   \sum _{j=1} ^{n-1} \langle n \vert H _n (\omega)\vert j \rangle 
   \ \langle j \vert \Psi (\omega) \rangle 
\,. \label {eq:411} \end {eqnarray}
The effective interaction in the  ($n -1$)-space becomes then 
\begin {equation}  
       H _{n -1} (\omega) =  H _n (\omega)
  +  H _n(\omega) G _ n  (\omega) H _n (\omega)
\label {eq:414} \end {equation}
for every block matrix  element 
$\langle i \vert H _{n-1}(\omega)\vert j \rangle$.  
Everything proceeds like in above,
including the fixed point equation  $ E  (\omega ) = \omega $.
But one has achieved much more: Eq.(\ref{eq:414}) is a 
{\em recursion relation} which holds for all $1<n<N$!
Since one has started from the bare Hamiltonian in the 
last sector, one has to convene that $H_{N}=H$. 
The rest is algebra and interpretation.

Applying the method to the block matrix structure of QCD, 
as displayed in Fig.~\ref{fig:holy-2}, is particularly easy and 
transparent.
By definition, the last sector contains only the diagonal kinetic 
energy, thus  $H_{13}=T_{13}$. Its resolvent is calculated trivially. 
Then $H_{12}$ can be constructed 
unambiguously, followed by $H_{11}$, and so on, until one 
arrives at sector 1. Grouping the so obtained results in a different 
order,  one finds for the sectors with one $q\bar q$-pair: 
\begin {eqnarray} 
     H_{q\bar q} = \qquad     H _1 
&=&  T_1 + V G _3 V + V G _3 V  G _2 V G _3 V 
\ , \label{eq:610}\\  
     H_{q\bar q\,g} = \qquad     H _3 
&=&  T_3 + V G _6 V + V G _6 V  G _5 V G _6 V + V G _4 V 
\ , \label{eq:620}  
\\  
     H_{q\bar q\,gg} = \qquad  H _6 
&=&   T_6 + V G _ {10} V+ V G _{10} V  G _9 V G _ {10} V 
                 + V G _7 V 
\,. \label {eq:622} \end {eqnarray}
The quark-gluon content of  the sectors is added here for easier 
identification, in line with the notation in Fig.~\ref{fig:holy-2}. 
Correspondingly, one obtains  for
the sectors with two $q\bar q $-pairs 
\begin {eqnarray} 
     H_{q\bar q q\bar q} = \qquad     H _4 
&=&  T_4 + V G _7 V + V G _7 V  G _6 V G _7 V  
\ , 
\\  
     H_{q\bar q q\bar q\, g} = \qquad     H _7 
&=&   T_7 + V G _ {11} V+ V G _{11} V  G _{10} V G _ {11} V 
                 + V G _8 V 
\,. 
\end {eqnarray}
In the pure gluon sectors, the structure is even simpler:
\begin {eqnarray} 
     H_{gg} = \qquad     H _2 
&=&  T_2 + V G _3 V + V G _5 V 
\ ,\label{eq:643} 
\\  
     H_{gg\, g} = \qquad     H _5 
&=&   T_5 + V G _ {6} V + V G _{9} V 
\ . \label{eq:645}\end {eqnarray}
Note that these relations are all exact. 

\begin{figure} [t]
\begin{minipage}[t]{50mm} 
\makebox[0mm]{}
\psfig{figure=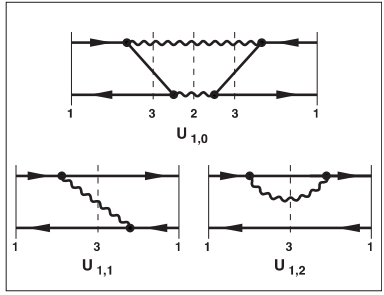,width=50mm} 
\vfill \end{minipage}
\hfill
\begin{minipage}[t]{60mm} \makebox[0mm]{}
\makebox[0mm]{}
\caption{\label{fig:6_1} 
  The three graphs of the effective interaction in the 
  $q\bar q$-space.~---
  The lower two graphs correspond to the chain $U=VG_3V$,
    the upper corresponds to $U_a=VG_3V G_2V G_3V$. 
   Propagator boxes are represented by vertical dashed lines, 
   with the subscript `$n$'  referring to the sector numbers. 
} \vfill \end{minipage}
\end{figure}

One is left with the eigenvalue problem in the $q\bar q$-space, 
\begin  {equation}
    \sum _{q^\prime,\bar q^\prime}
    \langle q;\bar q\vert H_{q\bar q} (\omega)
    \vert q^\prime;\bar q^\prime\rangle
    \psi_b(\omega) \rangle =  M_{b} ^2(\omega) 
    \langle q;\bar q\vert\psi_b(\omega) \rangle
\,.\label{eq:6.60}\end {equation} 
By construction, the eigenvalues are identical with those of the
full Hamiltonian and are enumerated by $b=1,2,\dots$. 
They are the invariant mass$^2$  of a physical particle and 
its intrinsic excitations.The  corresponding eigenfunctions 
$\langle q;\bar q\vert\psi_b(\omega) \rangle$  
represent the normalized projections of  $\vert\Psi\rangle$ onto 
the Fock states $ \vert q;\bar q\rangle 
   = b^\dagger _q d^\dagger_{\bar q} \vert vac \rangle $. 
The effective Hamiltonian as  given by Eq.(\ref{eq:610}) 
has two types of interactions which are illustrated 
diagrammatically in Fig.~\ref{fig:6_1}. In the first of them, in
$V G _3 V$, the bare vertex interaction scatters the system
virtually into the $q\bar q\,g$-space, where it propagates under
impact of the full interaction until a second vertex interaction 
scatters the system into the $q\bar q$-space. 
The gluon can be absorbed on the same quark line as it was 
emitted, which contributes to the effective quark mass as 
illustrated by diagram $U_{1,2}$ of Fig.~\ref{fig:6_1}.
The gluon absorbed by the other quark as in diagram $U_{1,1}$ 
of Fig.~\ref{fig:6_1} providing a quark-antiquark interaction which
cannot change the quark flavor.  It includes all fine and hyperfine 
interactions, see for instance  \cite{trp97a}. 
The second term in Eq.(\ref{eq:610}), 
the annihilation interaction $U_a=V G _3 V  G _2 V G _3 V$, 
potentially provides an interaction between different flavors
and is illustrated in diagram $U_{1,0}$ of Fig.~\ref{fig:6_1}. 
As a net result the interaction scatters a quark with helicity
$\lambda_q$ and four-momentum $p$
into a state with $\lambda_q^\prime$ and ${p^\prime}$.

The knowledge of $\psi_b$ is  sufficient to retrieve all desired 
Fock-space  components of the total wave-function. 
The key is the upwards recursion relation Eq.(\ref{eq:411}).
Obviously, one can express the higher Fock-space components 
$\langle n\vert\Psi\rangle$ as functionals of $\psi_{q\bar q}$ 
by a finite series of quadratures, {\it i.e.} of matrix multiplications 
or of momentum-space integrations.  
One need not solve an other eigenvalue problem.  
This is quickly shown by way of example, by calculating the 
probability amplitude for a  $\vert gg\rangle$- or a  
$\vert q\bar q\,g\rangle$-state 
in an eigenstate of the full Hamiltonian.
The first two equations of the recursive set in Eq.(\ref{eq:411}) are
\begin {eqnarray} 
        \langle 2 \vert\Psi\rangle &=& 
        G _ 2 \langle2\vert H_2\vert1\rangle 
        \langle 1 \vert \Psi  \rangle 
\,,\label{eq:448}\\  {\rm and}\quad
        \langle 3 \vert\Psi\rangle  &=& 
        G _ 3 \langle 3 \vert H _3 \vert 1 \rangle 
        \langle 1 \vert \Psi  \rangle +
        G _ 3 \langle 3 \vert H _3 \vert 2 \rangle 
        \langle 2 \vert \Psi  \rangle 
\,.\end{eqnarray}
The sector Hamiltonians $H _n$ have to be substituted from
Eqs.(\ref{eq:620}) and (\ref{eq:643}).
In taking block matrix elements of them, the formal expressions
are simplified considerably since most of the Hamiltonian blocks 
in Fig.~\ref{fig:holy-2} are zero matrices.  
One thus gets simply
$\langle2\vert H _2\vert1\rangle=\langle2\vert VG_3V\vert1\rangle$
and therefore
$\langle2\vert\Psi\rangle=G_2VG _3V \langle1\vert\Psi \rangle$.
Substituting this into Eq.(\ref{eq:448}) gives 
$\langle 3 \vert \Psi \rangle  = G _ 3 V\langle 1 \vert\Psi \rangle +
    G _ 3 VG _ 2 VG _3 V \langle1\vert\Psi\rangle$. 
These findings can be summarized more succinctly as
\begin {eqnarray} 
    \vert \psi_{gg} \rangle  &=& 
    G _ {gg} VG _{q\bar q\,g}V\:\vert\psi_{q\bar q}\rangle  , 
\\  {\rm and } \qquad
    \vert\psi_{q\bar q\,g}\rangle  &=& 
    G _ {q\bar q\,g} V\:\vert\psi _{q\bar q}\rangle +
    G _ {q\bar q\,g} VG _ {gg} VG _{q\bar q\,g} V 
    \:\vert\psi_{q\bar q}\rangle 
\ .\end{eqnarray}
The finite number of terms is in strong contrast to the infinite 
number of terms in perturbative series. Iterated resolvents sum 
the perturbative series to all orders in closed form.

\section{DISCUSSION AND PERSPECTIVES}

All of the above relations are exact and hold  
for an arbitrarily large $K$ \cite{pau96}.  
They hold thus also in the continuum limit, where the resolvents 
are replaced by propagators and the eigenvalue problems 
become integral equations. The effective interaction is 
very simple in direct consequence of the structure of the 
gauge field Hamiltonian with its many zero matrices. 
No particle cut-off is required, and no assumption is made on the
size of the coupling constant. 
The appoach is strictly non-perturbative.
Instead of the inverting a huge matrix as in the Tamm-Dancoff 
approach, one has to invert only the comparatively small 
sector Hamiltonians $H _n$.
The instantaneous interactions can be included easily ex post 
by the rule that every intrinsic line in graph in Fig.~\ref{fig:6_1} 
has to be suplemented with the corresponding instantaneous
line \cite{pau96}. 
Once the general structure of the effective interaction has been 
formulated, one can proceed with simplifying assumptions.
For example, one can replace the full by the free propagators 
to get a selected set of perturbative diagrams. The major effort 
of the ongoing work \cite{pau97} is to find an approximation 
scheme which combines rigor with simplicity. 
Unfortunately the limited space prevents giving more details.

\end{document}